\documentclass[aps,preprint,showpacs,preprintnumbers,amsmath,amssymb,floats,groupedaddress]{revtex4-1} 
\usepackage{graphicx}
\usepackage[english]{babel}
\usepackage{amsmath}
\usepackage{amssymb}
\usepackage{color}

\newcommand{\be}{\begin{eqnarray}} 
\newcommand{\ee}{\end{eqnarray}}

\begin{document}

\title{Theory of melting of glasses} 

\author{Chandra M. Varma}
\affiliation{Physics Department, University of California, Berkeley, Ca. 94704
}

\begin{abstract}
Glassy matter like crystals resists change in shape. Therefore a theory for their continuous  melting should show how the shear elastic constant $\mu$ goes to zero. Since viscosity is the long wave-length low frequency limit of shear correlations, the same theory should give phenomena like the Volger-Fulcher dependence of the viscosity on temperature near the transition.  A continuum model interrupted randomly by asymmetric rigid defects with orientational degrees of freedom is considered. Such defects are orthogonal to the continuum excitations, and are required to be imprisoned by rotational motion of the nearby atoms of the continuum. The defects interact with an angle dependent $\mu/r^3$ potential.  A renormalization group for the elastic constants, and the fugacity of the defects in 3D  is constructed. The principal results are that there is a scale-invariant reduction of $\mu$ as a function of length at any temperature $T < T_0$, above which it is 0 macrosopically but has a finite correlation length $\xi(T)$ which diverges as $T \to T_0$. Viscosity is shown to be  proportional to $\xi^2(T)$ and has the Vogel-Fulcher form. The specific heat is $\propto \xi^{-3}(T)$. As $T \to T_0$, the Kauzman temperature from above, the configuration entropy of the liquid is exhausted. The theory also gives the ``fragility" of glasses in terms of their $T_0/\mu$. 
\end{abstract}
\maketitle

Two of the principal mysteries of the glassy state are the nature of its melting  transition \cite{Angell1988, Jackle1986, Stillinger-rev, Ediger1996, Berthier-Biroli-RMP2011,  Bouchaud, Tarjus2005, Charbonneau} and the nature of the two level states
which appear to give the low temperature specific heat  $= \gamma T$ and other low temperature anomalies \cite{Pohl, PohlRMP2022, YuCarruzzo2021, AHV, Phillips1972}. Recently it has been found that
the coefficient $\gamma$ as well as the sound attenuation rate can be changed in two family of glasses by nearly two orders of magnitude by changing the quenching rate in preparation by vapor deposition \cite{Indo2014, Hellman2017}.  When $\gamma$ goes down, the thermodynamics near melting temperature $T_g$ develops a large peak \cite{Indo2014},\cite{CMV-JCCM2017} so that the  ground state entropy of the glass is reduced.  
 A relation, $\gamma ~T_g \approx constant$ \cite{Reynolds1980, Pohl1981} within classes of similar glasses had been noted earlier. These observations  suggest investigating a relationship between the two mysteries.  
 
\begin{figure}[h]
 \begin{center}
 \includegraphics[width= 0.8\columnwidth]{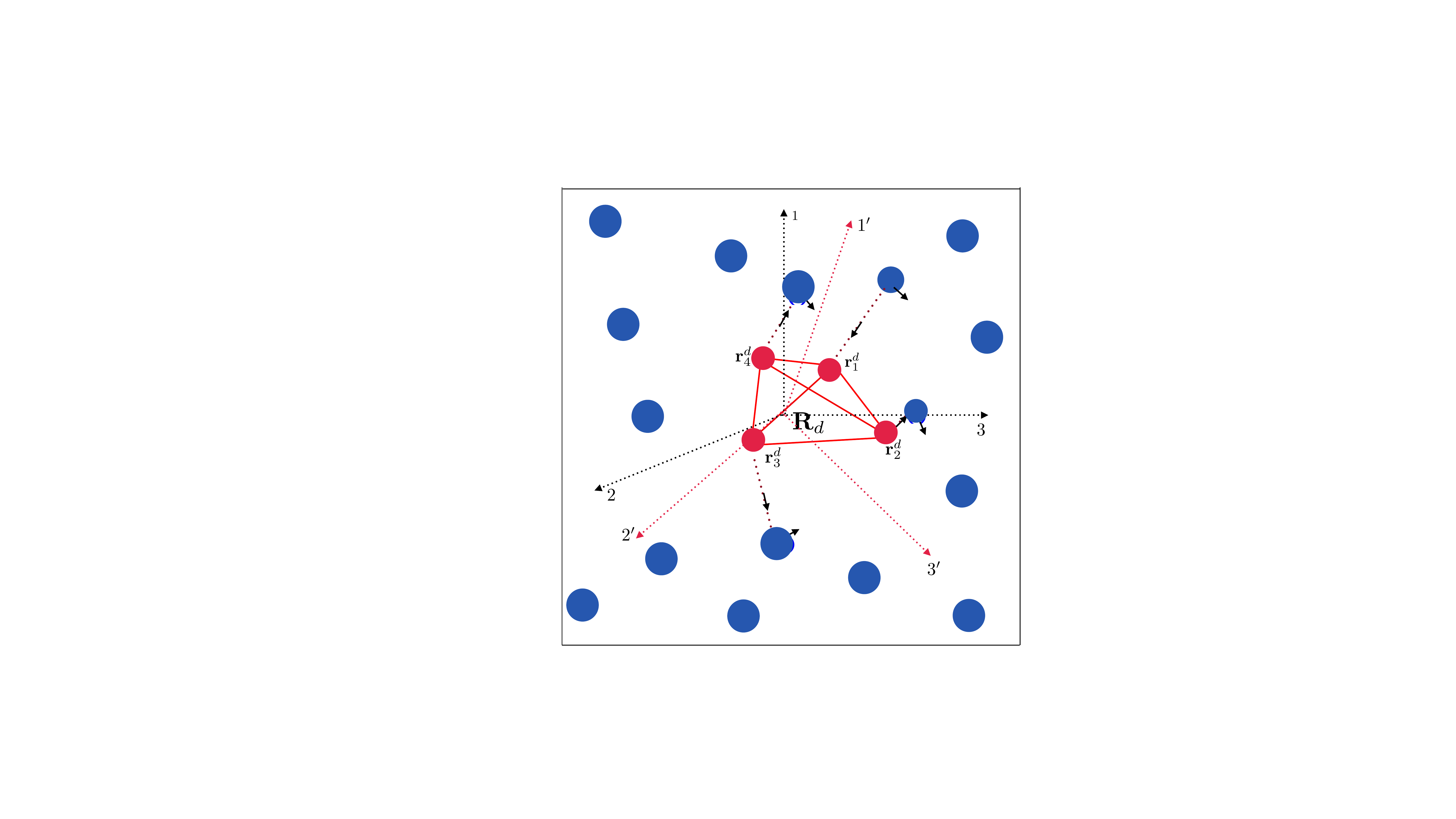}
 \end{center}
\caption{Simple model illustrating the geometry of the defects envisaged due to random atomic positions in a glass. A group of tightly bonded atoms (red) in the shape of a tetrahedra interacting weakly with its near neighbor atoms in the continuum is shown.  The orientation of such a rigid unit  is changed by rotation of the axes $(1,2,3)$, whose orientation is common to all defects, to $(1',2',3')$ accompanied, for force equilibrium and zeroing of torque, by the rotation and the displacements of the neighbors of the continuum.}
 \label{Fig:Defect}
\end{figure}

{\it The Model}: The question posed is, can there be a proliferation with temperature of configurational defects rather unique to the glassy state, which through their effects on the elastic correlation functions, change the shear constant of the glass to zero, provide the Vogel-Vulcher law in  transport properties and their observed relation to thermodynamics. Since the interaction between defects occurs due to long wave-length excitations, it is sufficient for purposes of this paper to consider a glass as an isotropic elastic media  interrupted randomly by a density of the defects at low temperatures. The defects are formed by small units on the scale of a few atoms with internal force constants much stronger than the coupling to their neighbors.  Due to their internal rigidity, such defects have rotational degrees of freedom. As discussed further below, a rigid rotation of a small unit must be accompanied by a compensatory rotation of atoms in the immediate vicinity. The re-defined defect is the rigid core surrounded by atoms in its immediate vicinity, which have the usual vibrational modes. The strain in the rest of the media can then be specified purely in terms of the stress on the surface atoms \cite{LL7, Clouet}. Fig. (\ref{Fig:Defect}) illustrates an orientable tetrahedral high density defect coupled with weak bonds to the neighbors. 

A rotation cannot be represented by shear (or compression) to any order in the lattice anharmonicity and is therefore orthogonal to the continuum excitations. It therefore fulfills a basic requirement for the two level states at low energy in glasses \cite{Lerner2021}. A small fraction of such states have low barriers between nearly equi-energy configurations to form the two level quantum-mechanical states. They can easily be incorporated. But our concern is with high temperature properties and we can ignore this aspect. 

The hypothesis of locally strong bonds accompanied by weak bonds is related to the idea that marginal stability or zero energy modes requires a balance between degrees of freedom and constraints \cite{PhillipsJC, Thorpe} and is applicable especially to theories of equilibrium and motion in granular media \cite{Nagle_Jamming}.  Constrained configurations elastically coupled have been used earlier to model \cite{Simon_V2001, He_C_V2010} the negative thermal expansion in a class of crystals. 
The idea is also related to the evidence that the liquid state prefers Franck configurations (icosahedra for example) with which periodic solids cannot be formed, and that glasses formed by a rapid quench may retain such configurations \cite{Nelson1983, Sethna1985, Tarjus2005}. Defects in models of glass discovered in numerical experiments \cite{Procaccia2007} in 2D may also be related. The defects are also closely related to the configurations for two level states found in numerical experiments \cite{Zamponi2020}. In numerical studies of silica glasses, rotational defects have been explicitly identified \cite{Heine2000}. In other numerical studies \cite{Lerner2021} defects with 
``Eshelby-like cores" surrounded by weak bonds to neighbors have been identified.   A specific ``Eshelby-like core" is defined in the model used here. 
  A simpler defect than shown in Fig. (\ref{Fig:Defect}) may be an atom with very strong bond attached at one end but very weakly coupled at the other elastically to the background. More complicated defects with similar properties depending on the chemical constituents of the model can be envisaged.  

A rigid defect  with $p = 1,..,\nu$ has corners at ${\bf r}^d_{p}$ given by a rotation ${\mathcal R}$ with respect to a chosen axes ${\hat e}^i, i =1,2,3$ with origin set in the given defect but orientation the same at all defects: 
\be
\label{rdp}
{r}^d_{pk} =    {\mathcal{R}_{ki}} b_{pi} {\hat e}^i.
\ee
(A small relative motion of the atoms of the polyhedra does not make any essential difference in the developments below.) It is accompanied by a a displacement ${\bf u}({\bf r})$    at the neighbors for local equilibrium. A local rotation without a corresponding change in the location of the atoms in the neighborhood in the continuum is not permitted in the solid state because it would lead to a  non-zero $\nabla \times {\bf u} ({\bf r})$ far from the defect and spoil the rotational invariance of the free-energy of the solid.  We wish to find relation of ${\mathcal R}$ and ${\bf u}({\bf r})$ to prevent that. Let us take the case that there is a spring constant only to the nearest neighbor in the continua. Let the neighbor  to the corner $p$ be displaced from its nominal position  at ${\bf R}_p = a_{pi}\hat{e}^i$ by ${\bf u}_p$. We must ensure the strains $u_{ij}$ and the rotations ${\mathcal R}$ are such that the sum of the forces on all the atoms in the defect (polyhedra plus its immediate neighbors in the continuum) and the torque on them is $0$. To do so, the potential energy of the defect is calculated.
 
The vector connecting the neighbor to the corner $p$ of the defect is specified in component form by
\be
({\bf \ell}_p^{i})_{k} =  a_{pi}(\delta_{ik}  + \partial_k u^i) - {\mathcal R}_{k i}b_{pi} \hat{e}^i.
\ee 
 The altered bond-distance $|\ell_p|$ from the corner $p$ to its nearest neighbor from the dot product is given by
$|\ell_p|^2 = (\ell_p^{i})_j(\ell_p^{i})_k.$
The change of the angle $\omega_p$ due to  ${\mathcal{R}}$ and $\nabla {\bf u}$
 is given by
\be
\label{angle}
& &\cos \omega_p = \frac{b_{pi}{\mathcal R}_{ki} (\ell_p^{i})_k}{|\ell_{p}|} 
= \frac{(b^2_{pi} -b_{pi} a_{pi}R_{ii}) + \frac{1}{2}{\mathcal{S}_{ik}}(\partial_i u_k  + \partial_k u_i) + \frac{1}{2}{\mathcal{A}_{ik}}(\partial_i u_k  - \partial_k u_i)}{|\ell_p|},
\ee
with $({\mathcal{S}_{ik}}, {\mathcal{A}_{ik}}) \equiv (a_{pi}b_{pk}R_{ik} \pm b_{pk}a_{pi}R_{ki})$.
 The potential energy in the defect is 
 \be
 \label{Pot}
 V =  \sum_{p} \frac{\mu_0}{2} \big( 1 - \cos\omega_p\big) + \frac{g_0}{2} \big(|\ell_{p}| - |a_p| \big)^2.
 \ee
 $\mu_0$ and $g_0$ are the shear and compressive elastic constants of the media. 
 The diagonal terms in the Euler angles ${\mathcal R}_{\alpha, \beta, \gamma}$ in Eq. (\ref{angle}) have correction quadratic in the Euler angles $(\alpha, \beta, \gamma)$,  while the off-diagonal terms are antisymmetric and linear in angles. To second order in the angles, the  condition that the derivative of energy with respect to angles be zero requires that $(\nabla \times {\bf u})$ is proportional to the Euler angles with a coefficient which depends on the local anisotropy ${\mathcal{A}_{ik}}$. The re-defined defect after this adjustment has no torque and far from it $\nabla \times {\bf u}({\bf r}) =0$.  The asymmetric stress due to the defect then presents a shear strain in the continuum, while the second term in (\ref{Pot}) presents a compressive strain. In matter with zero shear strain, i.e. the liquid, there would be no imprisonment of the  defect.
 
 There is a general way of describing local defects \cite{Clouet}, of which the particular defects discussed above are a special case.  Consider the closed surface of the defect.  The strain in the rest of the media can be calculated purely from the forces acting on the atoms at the  surface. One defines a second-rank tensor $P_{ij}$ with the dimensions of stress 
\be
P_{ij} = \frac{1}{v_d} \sum_n f^{(n)}_i r_j^{(n)},
\ee
where $n$ labels the atoms of the re-defined defect. $f^{(n)}_i$ is the force in the $i$-th direction on atom $n$, $\frac{\partial V}{\partial u_i^n}$ in Eq. (\ref{Pot}), and $r_j^{(n)}$ gives the locations of  $n$. $v_d$ is the volume of the defect. The sum of the forces on any atom in any direction $\sum_n f^n_i =0$. Also $P_{ij} = P_{ji}$ so that there is no torque.
This condition  ensures that far outside the surface of the defect,
$\nabla \times {\bf u}({\bf r}) =0$.  These conditions are equivalent to the imprisonment condition considered in the specific example above with the important fact that $P_{ij} \propto \mu_0$.

  {\it Renormalization of the elastic constants}:  The Green's function for displacement ${\bf u}({\bf r})$ for a continuum is given by
\be
G_{ij}({\bf r}) = \frac{1}{8 \pi \mu_0}\Big[A \delta_{ij} + B \eta_i \eta_j\Big]\frac{1}{|{\bf r}|}; ~~(A, B) \equiv 1 \pm \frac{\mu_0}{\lambda_0 + 2\mu_0}.
\ee
$\lambda_0$ and $\mu_0$ are the Lam{\'e} coefficients  which occur in the elastic constants as  
$C_{ijk\ell} = \lambda_0 \delta_{ij}\delta_{k\ell} + \mu_0(\delta_{ik}\delta_{j\ell} + \delta_{i\ell}\delta_{jk}),$
and $\eta_{i} = \frac{{\bf r}_i}{|{\bf r}|}$.
A defect with ${P}_{k \ell}$ at the origin produces a strain at a point ${\bf r}$ far from it \cite{Clouet}
\be
\label{streraain}
u_{ij}({\bf r}) =  - G_{i k,\ell j}({\bf r}) {P}_{k \ell}(0)
\ee
 $G_{i k,\ell j}(r)$ are the second derivatives in the directions $\hat{k}$ and $\hat{\ell}$ of  the  vector ${\bf r}$ in the Green's function $G_{i j}({\bf r})$.
The Fourier transform of the strain field for small ${\bf q}$ is
\be
\label{strain-q}
u_{ij}({\bf q}) = - \frac{1}{8 \pi \mu_0}\frac{q_j q_{\ell}}{q^2} \big( A \delta_{ik} + B \eta_i \eta_k\big) P_{k \ell}({\bf q}).
\ee
For  $P_{k \ell} \propto \mu_0$ the strains are dimensionless.
The interaction energy of a pair of defect $P_{k\ell}({\bf r})$  located at 
${\bf r}$ with a defect $P_{ij}(0)$ at the origin is
\be
\label{Eint}
E_{2} (P_{k \ell}({\bf r}), P_{ij}(0)) = P_{ij}(0) u_{ij}(0,{\bf r})
\ee
where $u_{ij}(0,{\bf r})$ is the strain at the location of the first defect due to 
$P_{k\ell}({\bf r})$. Let us define dimensionless stress at the defect $p_{ij} \equiv P_{i j}(8 \pi \mu_0)$. $E_2$ can be written as the interaction between the vectors $p_i({\bf r}) \equiv p_{ij} \eta_j({\bf r})$ and $p_{\ell}(0) \equiv \eta_k p_{k \ell} (0)$:
\be
\label{int}
\frac{E_2}{k_BT} ({\bf p}({\bf r}), {\bf p}(0)) = - 8 \pi \frac{\mu_0}{k_BT} \frac{1}{r^3} \big( A ~{\bf p}({\bf r}) \cdot  \bf p}(0) + B ~({\bf p}({\bf r}) \cdot {\hat{\bf r}) ({\bf p}(0) \cdot {\hat{\bf r}}) \big).
\ee
The fugacity of the pair includes also the probability of having pairs of non-interacting defects. We include the defect density at $T =0$ through $y_0^2$ and the density of thermally created defects (with cost in energy $2E_d$) to be proportional to $y_0^2$ with a coefficient $n'$ so that
\be
\label{fugacity}
y^2 = y_0^2(1+ n' e^{-2\frac{E_d}{k_BT}}).
\ee

It should be stressed that equilibrium statistical mechanical calculations are performed below. They are therefore not valid when equilibrium becomes 
impossible in measurements times near the transition. However the calculations give results valid as a theory of fluctuations of shear for temperatures on either side of the transition when equilibrium can be maintained. Most importantly, the theory will itself provide the time-required for equilibrium.

Introducing dimensionless elastic constants $(\overline{\mu}, \overline{\lambda}) = (\mu, \lambda)/(k_BT)$,  the change of the long-wave-length shear elastic constant $\overline{\mu}^R$ from $\overline{\mu}_0$ due to the shear strain produced by the defects is given by \cite{ChaikinLubensky, NelsonHalperin79}
\be
\label{renorm}
(\overline{\mu}^R)^{-1}_{ij} =(\overline{\mu}_0)^{-1}_{ij} + 4 k_BT \int_a^{\infty} d^3r  <u({\bf r}) u(0)>^D_{ij}
\ee
Here 
\be
 <u({\bf r}) u(0)>^D_{ij} \equiv <{u}_{ij}({\bf r}) {u}_{k\ell}({0})>^{D}(\delta_{i k}\delta_{j \ell} +
\delta_{i \ell}\delta_{j k})>.
\ee
is the correlation function of shear strains in the medium induced by the defects.
$a$ is the short distance cut-off given by the density.  
The strain-strain correlation $<uu>_{ij}^D$ induced by the defects can be expressed in terms of the defect correlations. Using the relation between $u_{ij}(q)$ in terms of $P_{k\ell}({\bf q})$ given by Eq. (\ref{strain-q}) and Fourier transforming,
\be
\label{corfn}
 < u({\bf r}) u(0)>^D_{ij}   
= \frac{1}{k_BT}\Big< \frac{1}{r^3}\Big(A^2 \big(\eta_i p_j +\eta_j  p_i)({\bf r})(\eta_j p_i +\eta_i p_j)(0)\big) + O(AB) +O(B^2)\Big)\Big>.
\ee
Two terms proportional to $AB$ and $B^2$ are not explicitly written down. But on calculating the expectation values as below, they contribute at most to $O(1/r^6)$ to the integrand after doing the angular integrals and are unimportant.

The expectation values are calculated  in the dilute limit, i.e.  to $O(y^2)$. This procedure is the same as  used  to calculate the renormalized shear correlation function due to the defects which were introduced in the classic works for the xy model for 2d superfluid and for 2d melting \cite{Kosterlitz1974, JKKN1977, NelsonHalperin79, Young79} \cite{ChaikinLubensky}. But the physical and mathematical features of our problem are quite different; we have  $1/r^3$ interactions in $d=3$ as opposed to the logarithmic interaction in $d=2$ in these works.  In the dilute limit $1/r^3$ interactions in $d=3$ gives results independent of the concentration of the defects. 
 
In calculating the renormalization of shear correlation, Eq. (\ref{corfn}), the relative direction of the vectors specified by the indices $i$ and $j$ are orthogonal and fixed. So the angular integrations in the expectation values are easy to do. The magnitudes of $|p_i|({\bf r})$ are assumed uniform. Then,
\be
\label{KR-s}
\overline{\mu}_R^{-1} = \overline{\mu}_0^{-1} + A^2y^2 \frac{\alpha}{32 \pi}\int_a^{\infty} \frac{dr}{r} 
e^{\alpha \overline{\mu}_R\frac{a^3}{r^3}},
\ee
with $\alpha = 16 \pi |p_i|^2$. Given this equation with its logarithmic singularity, the RG equations are obtained (details are given in a Supplementary Section) quite simply by replacing the upper limit in the integral by $R$ and calculating as a function of  $\ell = \ln (R/a)$ to get both $\overline{\mu}_R(R/a)$ and $y_R(R/a)$. This gives the the shear constant for $R \to \infty$ as well as the correlation length of shear. The RG equations obtained are,
 \be
 \label{RG1}
\frac{\partial y_R^2}{\partial \ell}  
&=& y_R^2 (1 - e^{\alpha \overline{\mu}_R}). \\
\label{RG2}
\frac{\partial \overline{\mu}_R^{-1}}{\partial \ell} &=& \frac{\alpha}{32 \pi} y_R^2 e^{\alpha \overline{\mu}_R}.
\ee
For $\alpha \overline{\mu}_R << 1$, Eqs. (\ref{RG1}) and  (\ref{RG2}) give
 \be
 \frac{\partial (y_R^2- 32 \pi \ln |\overline{\mu}_R|)}{\partial \ell} = 0.
 \ee
 The constant of the RG flow may be parametrized by $\ln(|\overline{\mu_c}|)$, so that
\be
\label{const}
y_R^2 =   (32 \pi) \ln\Big|\frac{\overline{\mu}_R}{\overline{\mu}_c}\Big|.
\ee
 Correspondingly,
\be
\label{muR}
\frac{\partial \overline{\mu}_R}{\partial \ell}  =  - \alpha \overline{\mu}_R^2 \ln\Big|\frac{\overline{\mu}_R}{\overline{\mu}_c}\Big|.
\ee
Let us first consider the glassy phase with $\mu_R \geqslant  |\mu_c|$ for all $\ell$.  Eqs. (\ref{const}, \ref{muR})  show a line of fixed points $\overline{\mu}_R = |{\overline{\mu}_c}|, y_R =0$. This line ends at the critical point for ${\overline{\mu}_c} =0$, where ${\overline{\mu}_R} = y_R =0$. In (\ref{muR}), $\mu_R$ continuously decreases with increasing $\ell$ for any $\mu_c$, and must stay at $0$ at $\ell \to \infty$ once that is a solution.  Evidently ${\overline{\mu}}_c \to 0$ tunes  the glass transition at temperature $T_0$. Returning to dimensional variables, we may put $\overline{\mu}_c  \approx \frac{\mu_0}{T_0}  (\frac{T-T_0}{T_0})$ close to $T_0$.

{\it Correlation length, viscosity and specific heat for $T >T_0$}: The continuous line of fixed points at $\mu = \mu_c, y =0$ for $T \leq T_0$ may be confirmed from the integration of Eq. (\ref{muR})  between $\ell(R/a)$ and $\ell(1)$ to get
\be
\label{soln}
 \ell i(|\mu_c/\mu(R/a)|) - \ell i(|\mu_c/\mu(1)|) = - (\alpha |\overline{\mu}_c|) \ln (R/a).
\ee
$\ell i(x) = P.V. \int_0^x dy/\ln y = \gamma + \ln \ln |x| + \sum_n \frac{(\ln |x|)^n}{n (n!)}$ \cite{A+S}. 
$\gamma$ is the Euler constant. For $|x|$ near 1, we need consider only the first two terms in $\ell i(x)$. With the boundary condition $\mu(1) = \mu_0$, Eq. (\ref{soln}) gives
\be
\label{muR-2}
\mu_R(R/a) = |\mu_c| \Big(|\frac{\mu_0}{\mu_c}|\Big)^{(\frac{a}{R})^{\alpha |\overline{\mu}_c|}}.
\ee
For asymptotically large $R/a$, the exponent is $0$ for any finite $|\mu_c|$ so that $\mu_R$ to $|\mu_c|$ with the approach to it slowing as $|\mu_c| \to 0$. The rate of approach allows the calculation of the correlation correlation length $(\xi/a)$ on either side of $T_0$ by expansion about $\mu = \mu_0$ at $R \to \infty$ for $T<<T_0$, and about $\mu =0$ in  liquid for $T << T_0$. The condition is that $\mu_R(\xi/a)$ as function of $|T-T_0|$ be below its glass value $\mu_0$ by some fraction, say $e$. Eq. (\ref{soln}) gives 
\be
\label{corlength}
\frac{\xi}{a} \approx e^{\frac{1}{\alpha |\overline{\mu}_c|}} = e^{(\frac{T_0}{\alpha \mu_0})\big(\frac{T_0}{T-T_0}\big)}.
\ee
In (\ref{corlength}) $(\overline{\mu}_c)\ln \ln (\mu_0/\mu_c)$ has been safely replaced by $\overline{\mu}_c$. The temperature dependence of the correlation length has the Vogel-Fulcher form. 

As already mentioned, this theory is an equilibrium theory for a {\it transition in a response function}, i.e. for long times compared to any relaxation rates. The free-energy change near the transition and therefore the specific heat $C_v \propto \big(\frac{a}{\xi}\big)^{3}$. The entropy for this specific heat is the configurational entropy of the liquid and it goes to $0$ as $T \to T_0$. This implies that $T_0$ is the Kauzman temperature, which is not achieved due to  very large relaxation times, which we calculate next.

The long relaxation times appear in measurements of the viscosity $\eta$ as well as the  specific heat \cite{Angell1988}. $\eta$ is the correlation (anti-commutator) of transverse stress $\sigma_{\bot}$ integrated over space and time \cite{KadanoffMartin1963}: 
\be
\label{eta}
\eta = \frac{1}{kT}Lim_{\omega \to 0, {\bf k} \to 0} \int d^3r \int_0^{\infty}  e^{i \omega t} e^{i {\bf k}\cdot {\bf r}}<\{\sigma_{\bot}(r,t), \sigma_{\bot}(0,0)\}>_T.
\ee
 For dense liquids, Maxwell gave the expression $\eta = \mu(L) \tau$, derived in Landau and Lifshitz \cite{LL7} - sec. 31. $\mu(L)$ is the shear over a length scale $L$ of molecular dimension and $\tau$ is the shear relaxation time.  Let us see how this comes about from (\ref{eta}). Divide the liquids into volumes whose shear are correlated, which in a liquid is given by $L$. If instead of integrating over $t$, we take the value for $t \to \infty$, (\ref{eta}) gives the static compressibility of the correlated region. But in the liquid, the correlation decays as $e^{-t/\tau}$, where $\tau$ is the rate of decay of the rotational or vibrational motion of the  molecules. The limit $\omega \to 0$ following the Fourier transform gives Maxwell's result. 
 
Near melting, we must divide the volume into shear correlated regions $(\xi/a)^3$, which are solid like and have shear compressibility of $O(\mu_0)$. The relaxation time $\tau$ is now given by shear to travel the correlated length $\xi$, which in random walk is of $\omega_s^{-1} (\xi/a)^2$, where $\omega_s$ is the frequency of local shear phonons. At $\xi =a$, the viscosity is that of the liquid, $\mu_0$, which is about $0.1$ poise for glycerol. Normalizing to it, one gets the Vogel-Fulcher law $\eta = \mu_0 (\xi/a)^2$. The ``fragility" of a specific glass  \cite{Angell1988, Angell_spht} is determined then by its $f \equiv \frac{2}{\alpha}(\frac{T_0}{\mu(a)})$, $\alpha  = 16 \pi |p_i|^2.$i  $p_i$ is expected also to be proportional to the shear elastic constant divided by temperature. A large glass transition temperature compared to the low temperature shear constant changes the Vogel-Vulcher law to the Arrhenius law. This is consistent with Glycerol being ``fragile" and Silica "strong".

 A thermodynamic "fragility" has been defined by Angell and by Debenedetti and Stillinger \cite{Angell_spht, Stillinger-rev, ShiraiSPHT}, and Angell has shown that the experimental data for the ``fragility" of thermodynamics and transport is the same and it is related the dependence of the two on the time of measurement. The specific heat falls dramatically as viscosity rises. These follow from above. These results bear correspondence with the heuristically motivated relation between specific heat and entropy by Adams and Dimarzio  \cite{Gibbs1958} long ago but the physical reasons in the derivation above are not the same. 

It has been widely surmised that there is a diverging correlation length
near the glass transition, \cite{Procaccia2007, Bouchaud, Nussinov2016}. The result here is quite natural, $\xi(T)$ is the length over which shear is correlated, going from molecular scale in the liquid to infinity in the solid state. It is also related to the density of locally mobile defects. This should be tested in further detail in numerical experiments but appears to be qualitatively consistent with Refs. \cite{Procaccia2007, Nussinov2016}, where a temperature dependence of the density of the defects has also been found. 

Surprisingly, very few measurements of the elastic constants near the transition appear to have been made. Some old measurements with a drop in the shear constant  to $0$ in glycerol which becomes more and more rapid as frequency of measurements decreases are given in Figs. (4 and 5) of Ref. \cite{Jackle1986}. A calculation of the compression modulus parallel to the shear
modulus given above gives for the former a similar decrease for all temperatures which must join to the modulus of the liquid for $T > T_0$. The remarkable non-analyticity in the result
for shear modulus in Eq. (\ref{muR-2}) is worth investigating carefully in experiments done by varying the wave-length and temperature, although with quite different techniques, a discontinuity is shown in  deductions  \cite{Keim2015} in colloidal glasses near melting.

 Many impressive but quite diverse directions  for the theory of glassy freezing have appeared since the clear systemization of data on glass formation by Kauzmann \cite{Kauzman1948}, Gibbs and DiMarzio \cite{Gibbs1958} and Angell \cite{Angell1988} very long ago.  The paper here draws on aspects of many of them which have been already mentioned.  An approach \cite{RFIM} which ties the freezing of glass to the theory of infinite range Ising spin-glass \cite{Charbonneau} with additional decorations leads to very successful predictions of non-linear susceptibilities \cite{Bouchaud}.  The ideas introduced here may be used to obtain non-linear susceptibilities and finite frequency properties of glasses by extending the theory to dispersion in the property of the defects using schemes such as the replica methods. The rotors identified here may be equivalent to the "spins" introduced in such theories. For $T$ significantly larger than $T_0$, the present theory with its liberated defects naturally goes over to the mode-coupling theories \cite{modecoupling}. The investigation of the possible liberation of the defects with non-linearity in applied stress may provide the underlying physics  for  the non-affine displacements found in the recent work on the theory and numerical simulation for plasticity in  glasses, see for example \cite{Zoccone2021}. Numerical studies of realistic models of glasses \cite{Reichman2022} are suggested to look further into the nature of defects that have been already found. 
 
Two ideas have been introduced in this work. One is asymmetric rigid rotor defects which are special to the glassy state and orthogonal to the usual excitations. Second is the use of the marginality of the
$1/r^3$ interactions between the dilute defects in 3d which makes a systematic RG theory possible. The defects of-course do not remain dilute as the transition is approached. Just as in the KT theory 
of Coulomb interactions between charges in 2d, this theory is expected to be consistent also at higher density of defects  because the renormalization of interactions of defects due to interposed additional defects does not essentially change the results for $1/r^3$ interactions, see for example \cite{Levitov1990}. 

{\it Acknowledgements}: 
  I thank Z. Nussinov and Clare Yu for discussions and comments on the manuscript. I also thank Frances Hellman for rekindling my interest in glasses This work was done partially as a visiting scholar at
 University of California, Berkeley; I wish to thank James Analytis, Robert Birgeneau and Joel Moore for making this possible. 
 
 \newpage
 \noindent
{\bf Supplement: Derivation of the Renormalization Group Equations \\
to ``Theory of melting of glasses"}\\

We start with Eq. (15) of the paper and  separate the integral into a part from $a$ to $ae^{\delta{\ell}}$ where $\ell = \ln (R/a)$. 
\be
\label{RGp}
\overline{\mu}_R^{-1} = \overline{\mu}_0^{-1} +  A^2 y^2 \frac{\alpha}{32 \pi} \Big(\int_a^{ae^{\delta \ell}} \frac{dr}{r} e^{\alpha \overline{\mu}_R \frac{1}{r^3}} + y^2 \int_{ae^{\delta \ell}}^{\infty} \frac{dr}{r}  e^{\alpha \overline{\mu}_R (\frac{a}{r})^3}\Big).
\ee
We first re-define the fugacity $y \to y_R$ to generate an equation in terms of renormalized quantities which looks identical to (15). This requires that  we change the limits of the second integral to be from $a$ to $\infty$ so that
 \be
 y_R^2\int_1^{\infty} \frac{dx}{x}e^{a\overline{\mu}_R \frac{1}{x^3}} = y^2 \int_{e^{\delta \ell}}^{\infty} \frac{dx}{x}e^{\alpha \overline{\mu}_R \frac{1}{x^3}}
 = y_R^2 \int_{1}^{\infty} \frac{dx'}{x'} e^{\alpha \overline{\mu}_R \frac{1}{x'^3}}(1- 3\alpha \overline{\mu}_R\frac{1}{x'^3} \delta \ell).
 \ee
 The integral  proportional to $\delta \ell$ can be calculated:
 \be
\int_1^{\infty} \frac{dx'}{x'^4} e^{\alpha \overline{\mu}_R \frac{1}{x'^3}} = \frac{1}{3} \alpha \overline{\mu}_R y_R^2 \big((1 - e^{\alpha \overline{\mu}_R}).
\ee
Therefore,
  \be
 \label{RG1}
\frac{\partial y_R^2}{\partial \ell}  
= y_R^2 (1 - e^{\alpha \overline{\mu}_R}).
\ee
The integral from $a$ to $a(1+ \delta \ell)$ of Eq.(\ref{RGp}) gives that
\be
\label{RG2}
\frac{\partial \overline{\mu}_R^{-1}}{\partial \ell} = \frac{\alpha}{32 \pi} y_R^2 e^{\alpha \overline{\mu}_R}.
\ee
In (\ref{RG2}) $A = 1 + \mu_R/(\lambda + 2 \mu_R)$ has been approximated by $1$, because $\lambda$ does not renormalize to $0$ (see below) and we are especially interested in the renormalization of $\mu_R$ towards $0$. Eqs. (\ref{RG1}) and (\ref{RG2}) are repeated in the text  
as Eqs. (16) and (17) to deduce the physical phenomena.

\newpage

%

\end{document}